\documentclass[journal=jpclcd,manuscript=letter]{achemso}
\usepackage{graphicx}
\usepackage{pgf}
\usepackage{amsmath,amssymb,lscape,mathtools}
\usepackage{color}
\usepackage{breqn}
\usepackage{alltt}
\usepackage{caption}
\usepackage{subcaption}
\usepackage{rotating}
\usepackage{tikz}
\usepackage{pdfpages}

\def\etal{\textit{et al.}~}
\DeclareSymbolFont{extraup}{U}{zavm}{m}{n}
\DeclareMathSymbol{\varheart}{\mathalpha}{extraup}{86}
\DeclareMathSymbol{\vardiamond}{\mathalpha}{extraup}{87}
\author{Abhishek Khetan}
\affiliation{Institute for Combustion Technology, RWTH, Aachen, Germany, 52056}
\author{Hamid R. Arjmandi}
\affiliation{Institute for Combustion Technology, RWTH, Aachen, Germany, 52056}
\author{Vikram Pande}
\affiliation{Department of Mechanical Engineering, Carnegie Mellon University, Pittsburgh, Pennsylvania, 15213}
\author{Heinz Pitsch}
\affiliation{Institute for Combustion Technology, RWTH, Aachen, Germany, 52056}
\author{Venkatasubramanian Viswanathan}
\affiliation{Department of Mechanical Engineering, Carnegie Mellon University, Pittsburgh, Pennsylvania, 15213}
\email{venkvis@cmu.edu}
\title{Understanding Ion Pairing in High Salt Concentration Electrolytes using Classical Molecular Dynamics Simulations and its Implications for Nonaqueous Li-O$_2$ Batteries}
\begin{document}
\begin{abstract}
A precise understanding of solvation is essential for rational search and design of electrolytes that can meet performance demands in Li-ion and beyond Li-ion batteries. In the context of Li-O$_2$ batteries, ion pairing is decisive in determining battery capacity via the solution mediated discharge mechanism without compromising heavily on electrolyte stability. We argue that models based on coordination numbers of the counterion in the first solvation shell are inadequate at describing the extent of ion pairing, especially at higher salt concentrations, and are often not consistent with experimental observations. In this study, we use classical molecular dynamics simulations for several salt anions (NO$_3^-$, BF$_4^-$, CF$_3$SO$_3^-$, (CF$_3$SO$_2$)$_2$N$^-$) and nonaqueous solvent (DMSO, DME, ACN, THF, DMA) combinations to improve the understanding of ion paring with the help of a new metric of ion-pairing. We proposed a metric that defines the degree of clustering of a cation by its counterions and solvent molecules on a continuous scale, the limits if which are based on a simple and intuitive condition of charge neutrality. Using these metrics, we identify the extent of ion pairing in good agreement with experimental solvation phase diagrams and further discuss its usefulness in understanding commonly employed measures of salt and solvent donicity such as the Gutmann donor number. \\
\begin{tocentry}
	\includegraphics[width=3.0in]{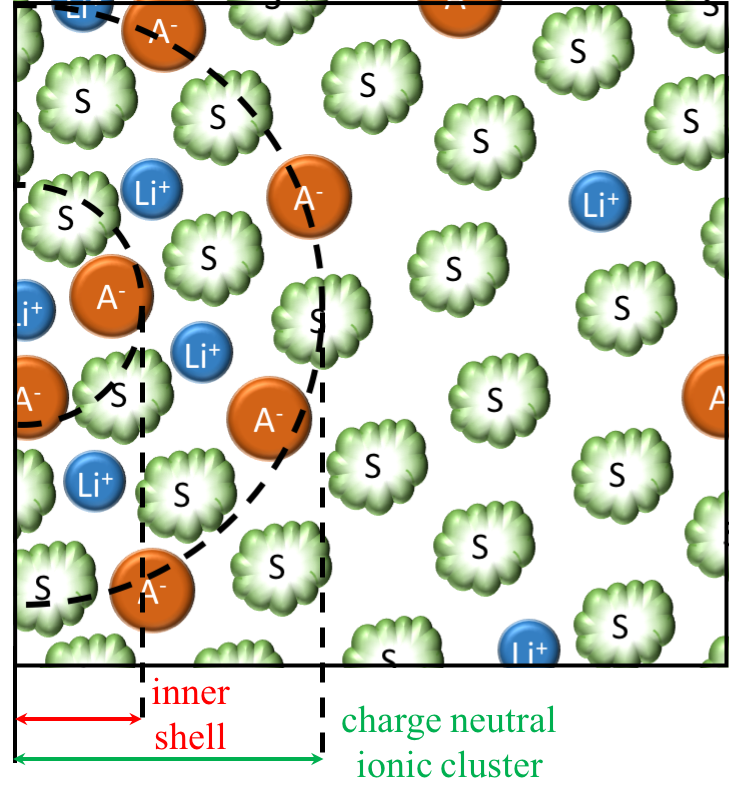}
\end{tocentry}
\end{abstract}
\newpage

\section{Introduction}
Despite having the highest theoretical specific energy among ``beyond Li-ion'' (BLI) battery chemistries, nonaqueous Li–O$_2$ batteries are yet to make significant technological strides toward commercialization due to several critical challenges such as low rechargeability and capacity~\cite{McCloskeyLuntz2014,Oleg2015,BruceAurbach2016}. The electronically insulating nature of the primary discharge product Li$_2$O$_2$, which passivates the battery cathode as it is formed, puts fundamental limitations on the extent to which discharge reactions can take place at a feasible rate. Recently, strategies to circumvent this issue by enhancing solubility of the reacting intermediates have been reported, which are based upon electrolyte formulations that use solvents~\cite{Trahan2014,BruceJohnson2014} with high Gutmann Donor Number (DN) and Acceptor Number (AN) for stabilizing the primary reacting ions Li$^+$ and O$_2^-$ and their solvated intermediates such as LiO$_2$ and LiO$_2^-$. However, it has been shown using thermodynamic analysis that aprotic solvents' ability to induce the solubility of these intermediates and thus provide for higher discharge capacity is anti-correlated to their stability against nucleophilic attacks~\cite{Khetan:2015aa}, H-abstraction~\cite{Khetan:2014ab} and electrochemical oxidation~\cite{Khetan:2014aa}. In this regard, the effect of Li-salts~\cite{Nasybulin2013,Gunasekara2015,AurbachSharon2016} and additives~\cite{Aetukuri2015,BruceGao2016} has been also explored to in order to influence the solubility of the intermediates without compromising on the stability of the solvent, with reports of multifold capacity enhancement. For instance, the capacity enhancing effect of NO$_3^-$ anions in comparison to TFSI has been reported widely in literature but uncertainities exist owing to several different explanations for this phenomemon, ranging from higher donicity and ionic association strength of the NO$_3^-$ anion~\cite{Burke2015,AurbachSharon2016} to its catalytic effect on the electrodes~\cite{Walker:2013aa,AurbachSharon2015} to the deactivation of the usually employed carbon cathode to further degradation~\cite{KimKang2014,SauerIliksu2017}. 

Using mixtures of two commonly employed Li-salts, LiNO$_3^-$ and LiTFSI, dissolved in dimethoxyethane (DME), we demonstrated in an earlier work that electrolytes containing a high NO$_3^-$:TFSI ratio resulted in higher discharge capacity~\cite{Burke2015}. Based on the principle that the free energy of the Li$^+$ ions and thus, the free energy of dissolution of the adsorbed intermediate LiO$_2$*, is largely dependent on the species that are present in the first solvation shell, we developed a modified Ising model for site occupancy in the shell. The estimated fractional occupation numbers resulting from this work provided a rational basis for selection of the total electrolyte, i.e., solvent and anion, where it was also shown that here is minimal capacity enhancement by changing the electrolyte anion in high-DN solvents. The modified Ising model served a useful tool for phenomenological understanding, however a more refined picture is required to estimate the solvation structure and dynamics to cover the broad spectrum of behavior spawned by varying salt-solvent combinations. 

The reactivity, selectivity and thermophysical properties of electrolytes depend strongly on the solvation behavior of the reactant ions, which is described by the degree of association of oppositely charged ions in electrolyte solutions to form distinct chemical species called ion pairs~\cite{Marcus2006}. It has been shown that solvation and ion pairing behavior has far reaching implications on the electrochemical and chemical stability of both the nonaqueous solvent and the salt anion, which is intricately linked with the coupling between them~\cite{AmineDu2013,PerssonRajput2015}, and which cannot be described by simple atom centered descriptors such as Mulliken charges or electron affinity~\cite{Bryantsev:2013ab}. First order descriptors such as DNs have been successful in explaining trends in capacity and stability but are too coarse to be able to provide reliable trends for use in rational electrolyte design and discovery~\cite{Oleg2015}. 

\begin{figure}[!ht]
\centering
\captionsetup{format=plain}
\includegraphics[width=1.0\textwidth]{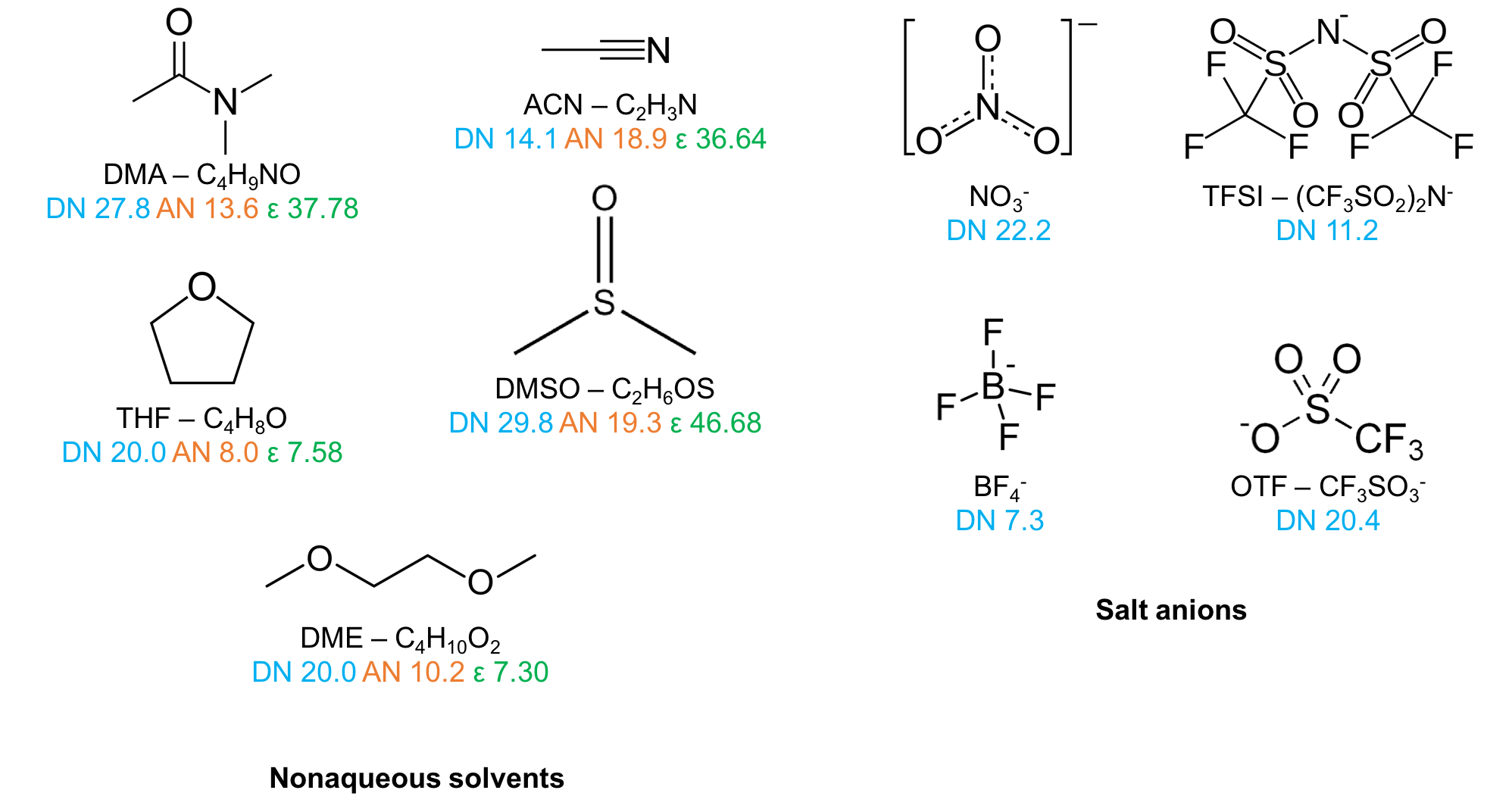}
\caption{Nonaqueous solvents and salt anions considered in the current work with their Gutmann Donor numbers (DN), Acceptor numbers (AN) and dielectric constants($\epsilon$)~\cite{Gutmann:1976aa,Mayer:1979aa,Schmeisser:2012}.}
\label{fig:ss}
\end{figure}

There is much subjectivity as to what forms an ion pair and it is often inferred from spectroscopic experiments or based on a certain distance of separation, $r$, between two oppositely charged ions in solution, which is smaller than some specified cutoff distance, $R$, beyond which ions are considered free~\cite{Marcus2006}. Electrolytes in which ionic interactions become prominent, also termed often as``solvates''~\cite{WatanabeUeno2012} remain difficult to characterize, as current theories cannot model their attributes adequately.  In this work we make use of rigorous classical molecular dynamics (MD) simulations to provide a general description of the solvation structure of Li$^+$ ions and the role of both solvent and anion in ion-pairing for several combinations of nonaqueous solvents and salt anions over a wide range of electron donicities and sizes, as shown in Fig.~\ref{fig:ss}. Classical MD simulations help us take into account the effect of size, the averaged charge distribution within a molecule or ion, and their correlation with the intermolecular interactions and the resulting solvation structure.

Extensive studies of the solvation behavior of Li$^+$ ions using experiments and classical MD simulations for several different salt anions have been performed in solvent like acetonitrile (ACN)~\cite{HendersonSeo2012a,HendersonSeo2012b,HendersonSeo2013,HendersonHan2013,HendersonHan2014}, dimethyl sulfoxide (DMSO)~\cite{AkagiJung2016}, carbonates~\cite{SeminarioKumar2016} and glymes of different chain lengths~\cite{ChupasLapidus2014,WatanabeTsuzuki2015,WatanabeSaito2016,NatanKuritz2016}. However, computational descriptions of ion pairing typically consider ion-pairing from the solvation structure in the first solvation shell~\cite{Henderson2006,PerssonRajput2015}, and it can be shown that the conventional choice of the radius defining this shell has limited applicability and cannot account for the extent of aggregation and clustering size.  In this work we propose a novel way to quantify the extent of ion-paring by defining a simple and intuitive metric based on charge neutrality of ionic clusters. Using this metric we offer a new generalized classification scheme for ion pairing and compare the various salt solvent combinations considered in this work. Additionally, we discuss the applicability of the commonly employed donor number in understanding solvation. This proposed method as such will be independent of the accuracy of forcefields or level of theory used for computation.

\section{Computational Details}
We make use of the GROMACS code v.5.1.2~\cite{LindahlAbraham2015} and OPLS-AA forcefields~\cite{CalemanSpoel2012,SpoelCaleman2012}. The forcefields for anions X$^-$ =  NO$_3^-$, BF$^-_4$, CF$_3$SO$^-_3$ (OTF) anions was taken from the work of \textit{Acevedo}~\etal~\cite{AcevedoRao2009}, for (CF$_3$SO$_2$)$_2$N$^-$ (TFSI) anion was taken from the work of \textit{Lopes and Padua}~\etal~\cite{PaduaLopes2004}, for DME was taken from the work of \textit{Watanabe}~\etal~\cite{WatanabeTsuzuki2015}. The forcefields for all other molecules were taken directly from the OPLS-AA forcefield repository~\cite{CalemanSpoel2012,SpoelCaleman2012}. The used forcefield parameters, topologies and assigned partial charges can be found in the Supplementary Information

All simulation boxes were chosen to be  5.5 $\times$ 5.5 $\times$ 5.5 nm$^3$ in size with periodic boundary conditions such that 100 Li$^+$ ions and salt anions in the box resulted in a total salt concentration of 1 M. After equilibration for 5 ns in an NVT ensemble and subsequent 10 ns of equilibration in NPT ensemble, the production runs for the simulations were performed in for a further 12 ns. The Parinello-Rahman barostat was used to maintain the system at a pressure of 1 bar and the N{\'o}se-Hoover thermostat was employed to maintain the system at 298.15 K. The Particle-Mesh Ewald (PME) method was used to consider long-range electrostatic interactions, which were truncated at 1.3 nm along with those resulting from non-bonded Lennard-Jones interactions. All the analysis results were averaged over a couple of independent configurations of the same system. In addition to the all the solvent and salt combinations shown in Fig.~\ref{fig:ss}, mixtures of  NO$_3^-$ and TFSI of total 1 M concentration were also simulated in order to understand the competition between NO$_3^-$, which is a small, strongly associating anion to TFSI, which is a large, weakly associating anion.

\section{Results and Discussions}
In order to gauge the ordering of ions and solvent molecules around the Li$^+$ ions, we first discuss and plot the time averaged normalized radial distribution functions (RDF) of all electrolyte constituents for the cases of 1 M LiNO$^-_3$ and 1 M LiTFSI in ACN, DME and DMSO in Fig.~\ref{fig:rdfset}. The RDFs for other salt-solvent combinations considered in this work can be found in the Supplementary Information. The RDF or pair correlation function of species $j$ around species $i$ as a function of radius $r$ is defined as:

\begin{equation}\label{eq:rdf}
g_{ij}(r) = \frac{\langle \rho_{j}(r) \rangle}{\langle \rho_{j} \rangle_{avg}}
\end{equation}
where the term $\langle \rho_{j}(r) \rangle$ is the density of $j$ within an infinitesimal region at $r$ and $\langle \rho_{j} \rangle_{avg}$ is the averaged local density of $j$ in the volume contained by $r$. The variations in $g_{ij}(r)$ signify the relative probability of finding $j$ at a distance $r$ from the reference $i$, for e.g., a the closest peak in $g_{ij}(r)$ indicates a particularly favored separation distance where $j$ is likely to be found around $i$. The magnitude of $g_{ij}(r)$ is a function of binning size for $r$ and its absolute value cannot be used to make quantitative comparisons directly. In macroscopic homogeneous systems, $g_{ij}(r)$ goes to unity at large $r$. In the present work, $g_{ij}(r)$ was plotted after normalization with respect to its maximum value for purposes of representation. 

\begin{figure}[!ht]
\centering
\captionsetup{format=plain}
\includegraphics[width=1.0\textwidth]{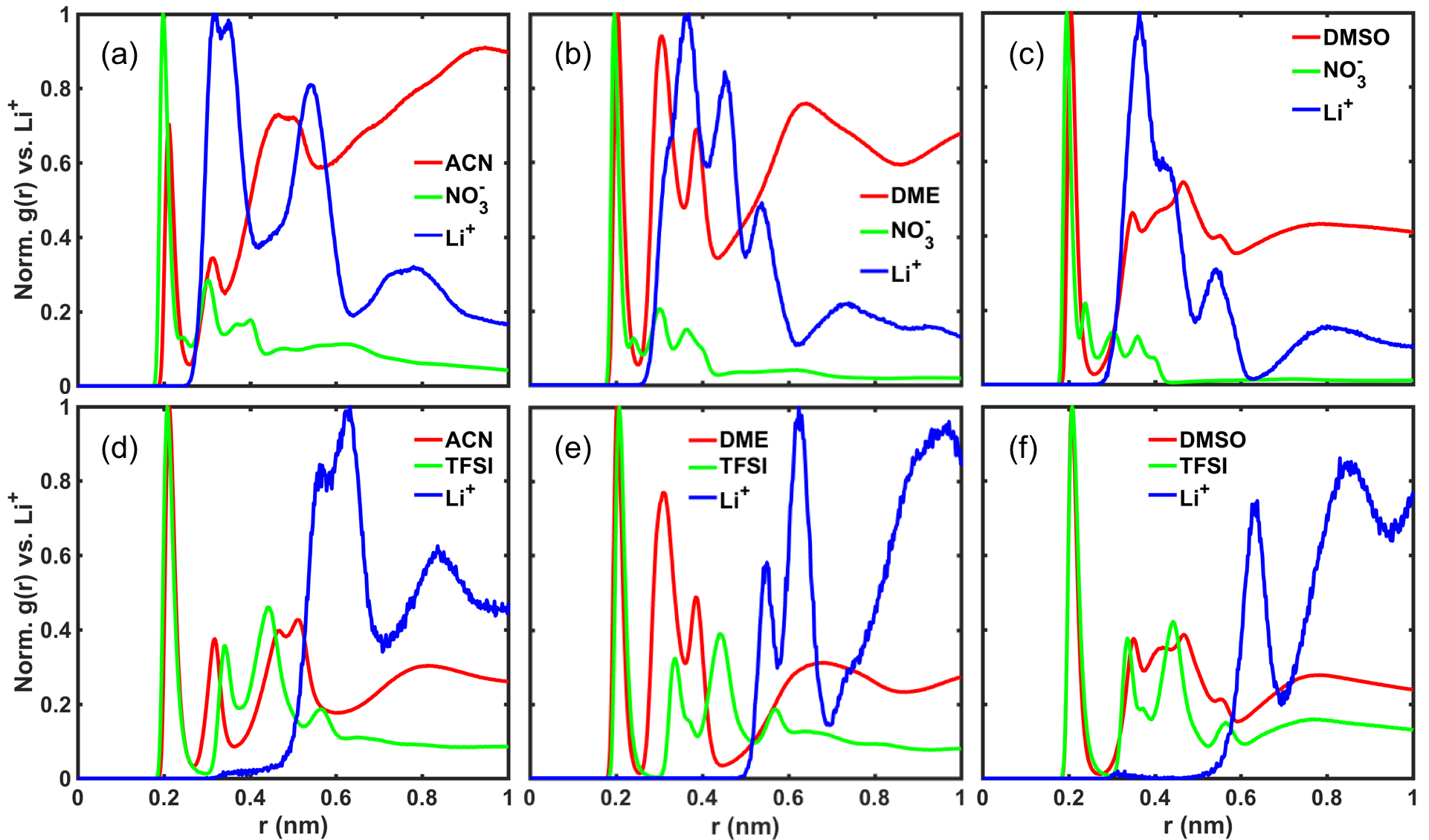}
\caption{Normalized RDFs of electrolyte components around Li$^+$ ions for (a) ACN - LiNO$_3$, (b) DME - LiNO$_3$, (c) DMSO - LiNO$_3$, (d) ACN - LiTFSI, (e) DME - LiTFSI, (f) DMSO - LiTFSI.}
\label{fig:rdfset}
\end{figure}

On comparing Fig.~\ref{fig:rdfset}a, Fig.~\ref{fig:rdfset}b and Fig.~\ref{fig:rdfset}c with NO$_3^-$ as anion across the three different solvents, it can be observed that the nearest of all maxima always belonged to the NO$_3^-$ anion at $r$ = 0.198 nm (ACN), 0.196 nm (DME, DMSO) followed very closely by that solvent $\sim$ 0.01 nm thereafter. This behavior clearly demonstrates that NO$_3^-$, which is known to be strongly associating, dominates the immediate solvation surroundings of the Li$^+$ ions irrespective of the solvent. The radial distance marking the first minimum in the RDF, or alternately the radius of the inner solvation shell, is found to be at  $r_{sh}^{NO^-_3}$ = 0.236 nm (ACN), 0.228 nm (DME), 0.226 nm (DMSO). On comparing Fig.~\ref{fig:rdfset}d, Fig.~\ref{fig:rdfset}e and Fig.~\ref{fig:rdfset}f with TFSI as anion across the three different solvents, it can be observed that the first maximum belonged to TFSI only in the case of ACN at $r$ = 0.208 and the first solvent maxima preceded that of the TFSI anion in cases of DME and DMSO. This behavior demonstrates that for the weakly associating TFSI anion the high donor solvents can play an equal or dominant role in the liquid ordering around Li$^+$ ions. The radius of the inner solvation shell in these cases was found to be  $r_{sh}^{TFSI}$  = 0.298 nm (ACN), 0.292 nm (DME), 0.294 nm (DMSO). 

\begin{figure}[!ht]
\centering
\captionsetup{format=plain}
\includegraphics[width=1.0\textwidth]{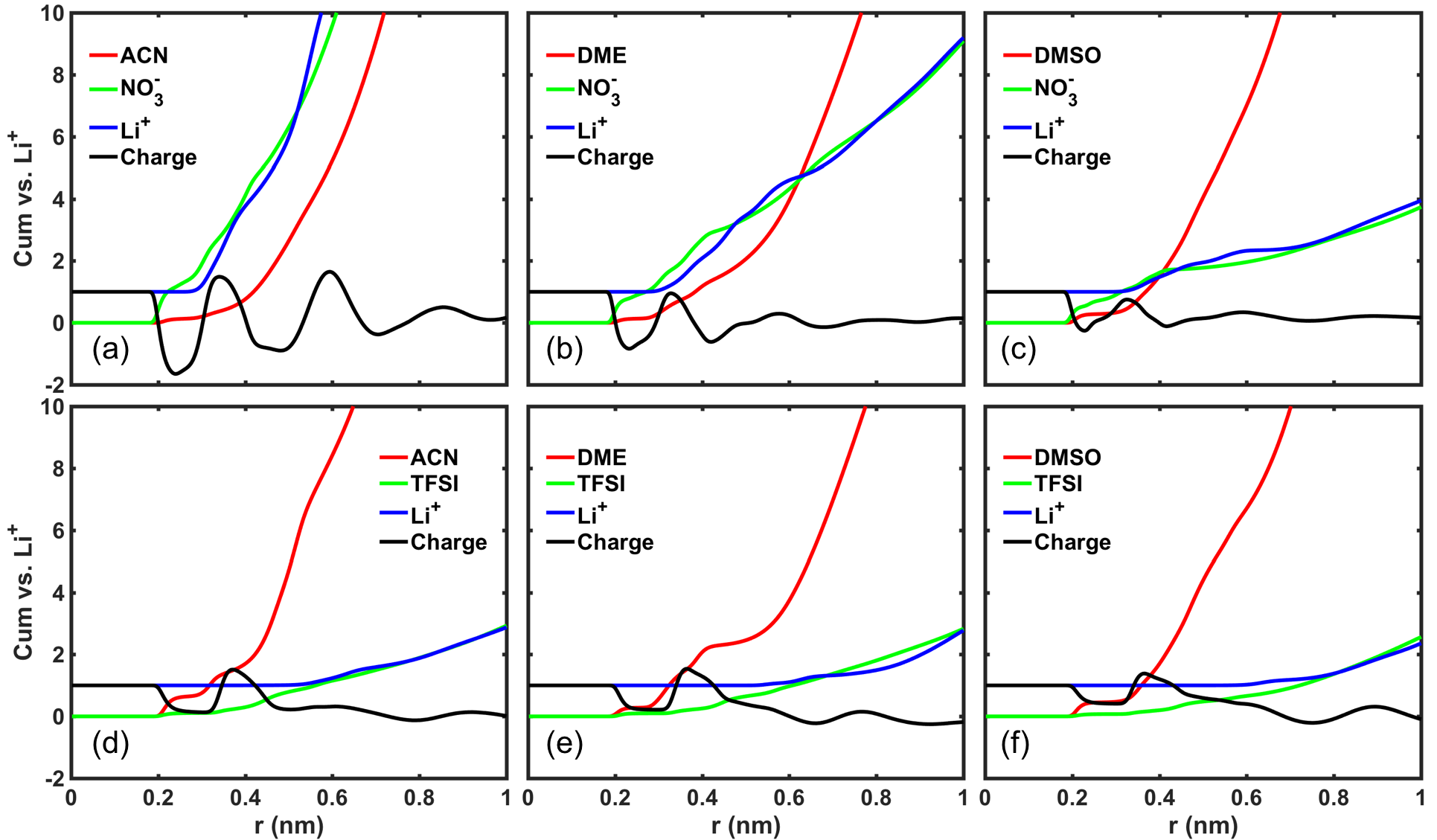}
\caption{Cumulative numbers for various electrolyte constituents around Li$^+$ ions and the charge contained within radius $r$ for (a) ACN - LiNO$_3$, (b) DME - LiNO$_3$, (c) DMSO - LiNO$_3$, (d) ACN - LiTFSI, (e) DME - LiTFSI, (f) DMSO - LiTFSI.}
\label{fig:cumset}
\end{figure}

The RDFs do not directly reveal the occupancy or coordination numbers (CN) of electrolyte constituents around any Li$^+$ ion. Cumulative values of each coordinating specie can are obtained by the integration of the function $g_{ij}(r)$, as shown in Fig.~\ref{fig:cumset} for the cases of 1 M LiNO$_3$ and 1 M LiTFSI in ACN, DME and DMSO. The CNs are then simply the cumulative values of the respective components at the respective inner shell radii $r_{sh}$. In line with the RDFs, the cumulative values show a much larger presence of the NO$^-_3$ anion closer to the Li$^+$ ions, which decreases as the DN of the solvent increases in the order ACN $<$ DME $<$ DMSO. This is again in stark contrast to the case of TFSI anions where the solvents dominate in presence around the Li$^+$ ions and there is not much difference in the calculated cumulative values. In addition to the cumulative numbers of the electrolytes constituents, the total charge $q_{r}$ contained in the radius $r$ around any Li$^+$ ion is also shown, which is evaluated simply as the algebraic sum of the forcefield assigned atomic charges belonging to the atoms in the solvating cations and anions. $q_{r}$ = 1 at $r$ = 0 because of the reference Li$^+$ ion. Including the atomic charges from solvent molecules served only to smooth the plotted charge curves and was therefore not pursued further. 

\begin{figure}[!ht]
\centering
\captionsetup{format=plain}
\includegraphics[width=0.8\textwidth]{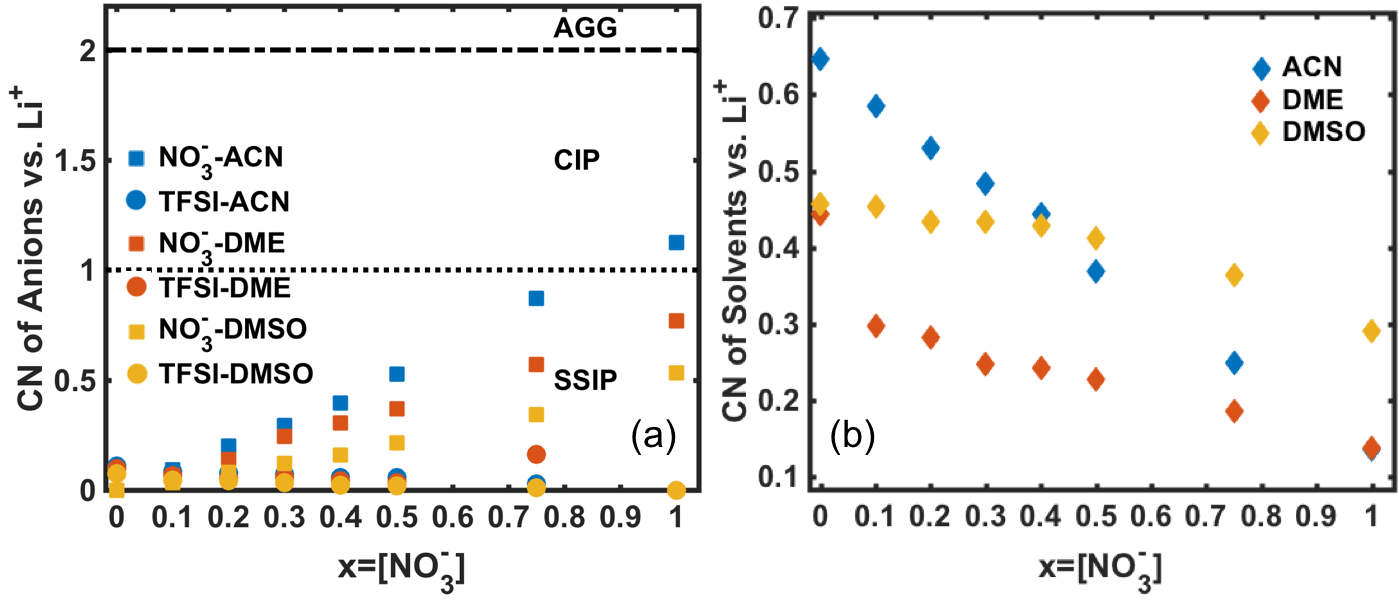}
\caption{CNs of electrolyte components in the inner solvation shell of Li$^+$ ions as a function of electrolyte composition for (a) anions and (b) solvents.}
\label{fig:CNs}
\end{figure}

The standard method of classifying ion pair formation as solvent separated (SSIP), contact (CIP) or aggregating (AGG) ion pairs is based on the coordination number (CN) of the counterion in the first solvation shell being between 0 to 1, 1 to 2 or more than 2~\cite{Henderson2006,PerssonRajput2015} . To examine the applicability of this scheme, we first plot the CNs of the anions and the solvents ACN, DME and DMSO in the inner solvation shell of Li$^+$ ions as a function of varying NO$^-_3$:TFSI ratio in the electrolyte composition, as seen in Fig.~\ref{fig:CNs}. The cumulative values and CNs for other salt-solvent combinations considered in this work can be found in the Supplementary Information.  As can be seen in Fig.~\ref{fig:CNs}a, the CN of NO$_3^-$ in the inner solvation shell at $r_{sh}$ increases linearly with increase in NO$_3^-$ concentration and is an order of magnitude higher than the CN of TFSI which falls to 0 with increasing NO$_3^-$ concentration. For both anions, the CNs decrease as the donicity of the solvent increases from ACN $<$ DME $<$ DMSO. The donicity of solvents or anions alone, however, does not decide its CN in the inner solvation shell. As can be seen in Fig.~\ref{fig:CNs}b, the CN of weakly donating ACN at lower NO$_3^-$ concentrations is higher than those of DME or DMSO, indicating that size effects as well as the anion - solvent compatibility may also play a role. 

Based on this previosuly employed classification scheme, CIP formation can be inferred only for the cases where [NO$^-_3$] $>$ 0.75 M in ACN. This assertion, however, is not consistent with room temperature experimental phase diagrams in which NO$^-_3$, CF$_3$SO$^-_3$ are already insoluble in ACN~\cite{HendersonSeo2012a}, whereas TFSI, BF$^-_4$  form SSIP/CIP in ACN~\cite{HendersonSeo2012a,HendersonSeo2012b}, at ACN:LiX ratio of 1:20 (considered in this work). While the CNs do reveal the fractional ordering in the immediate solvation shell averaged over every Li$^+$ ion and over time, they do not reveal the extent of ion pairing and formation of aggregates. Besides, the definition of CN and its associated $r_{sh}$ implies that CN of Li$^+$ vs. Li$^+$ is always 0, and hence, they do not contain any information about the self-distribution of Li$^+$ ions around itself. Therefore, classifications of ion pairs as SSIP, CIP or AGG are rather restricted to strict and limited models which do not include, for instance, clusters of ion pairs with infused solvent molecules, where counterions may not aggregate without necessarily being in direct contact with each other.

\begin{figure}[!ht]
\centering
\captionsetup{format=plain}
\includegraphics[width=1.0\textwidth]{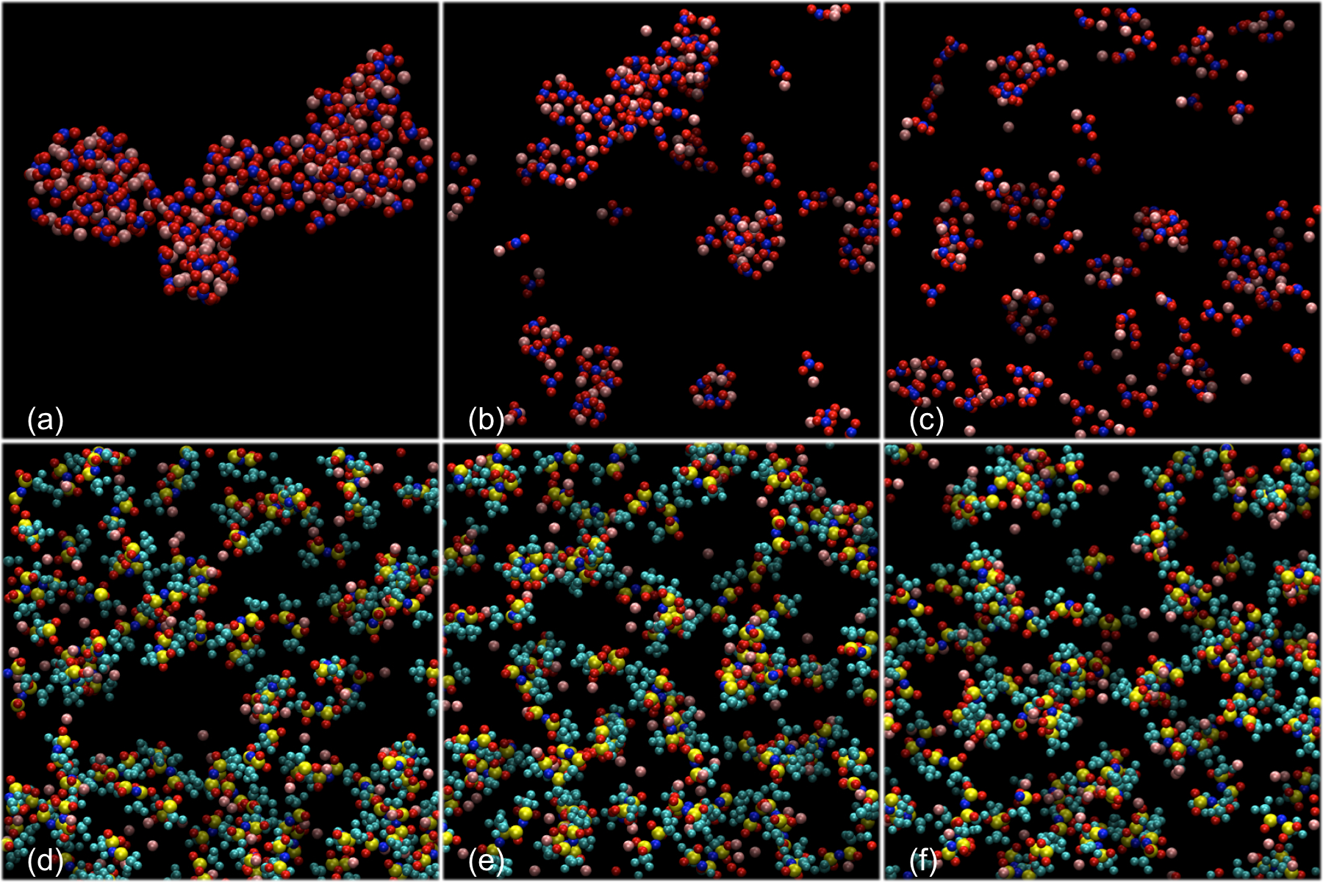}
\caption{Representative snapshots of simulation boxes shown without solvents for (a) ACN - LiNO$_3$, (b) DME - LiNO$_3$, (c) DMSO - LiNO$_3$, (d) ACN - LiTFSI, (e) DME - LiTFSI, (f) DMSO - LiTFSI. The atoms are colored as Li (pink), N (blue), O (red), F (cyan) and S (yellow).}
\label{fig:boxset}
\end{figure}

The inconsistency pointed out here becomes more evident on analyzing the snapshots from MD simulations in each case, as shown in Fig.~\ref{fig:boxset}. Several snapshots were taken from every simulation run to select the most representative insights about ion pairing for each of the systems. In every snapshot, we take out the solvent molecules and show the ions only in order to focus on ion-pairing between Li$^+$ ions and the salt anion. When comparing the snapshots from Fig.~\ref{fig:boxset}a, Fig.~\ref{fig:boxset}b and Fig.~\ref{fig:boxset}c with 1 M NO$_3^-$ anion across the three different solvents, it can be observed that there is a high degree of aggregation in the low DN solvent ACN, which decreases rapidly as the donicity of the solvent increases. On the contrary, not much perceivable difference is observed for when comparing Fig.~\ref{fig:boxset}d, Fig.~\ref{fig:boxset}e and Fig.~\ref{fig:boxset}f, with the weakly associating TFSI as anion across the three different solvents. These observations, however, cannot be deduced using the simple scheme of using CNs for ion pairing classification as has been performed in previous studies.

In order to address this discrepancy, we propose a new metric for understanding ion pairing based on the more general and intuitive condition of charge neutrality. Any distinct cluster of ions and solvents will naturally be charge neutral within a certain tolerance. An analysis of the radial charge $q_{r}$ contained within $r$, as seen in Fig.~\ref{fig:cumset}, reveals the charge distribution of the species dominating in presence in the cluster depending of the sign of the charge curve. Clearly with NO$^-_3$ as the anion, the fluctuation in the charge is much higher in the low DN ACN. These fluctuations damp down as the DN of the solvent increases from ACN to DME to DMSO. The same is corroborated from the snapshot in Fig.~\ref{fig:boxset} where the degree of clustering can be seen to drastically decrease. In case of the TFSI anions, the snapshots hardly show any variation in ion pairing across the solvents as do the charge curves from Fig.~\ref{fig:cumset}.

\begin{figure}[!ht]
\centering
\captionsetup{format=plain}
\includegraphics[width=0.75\textwidth]{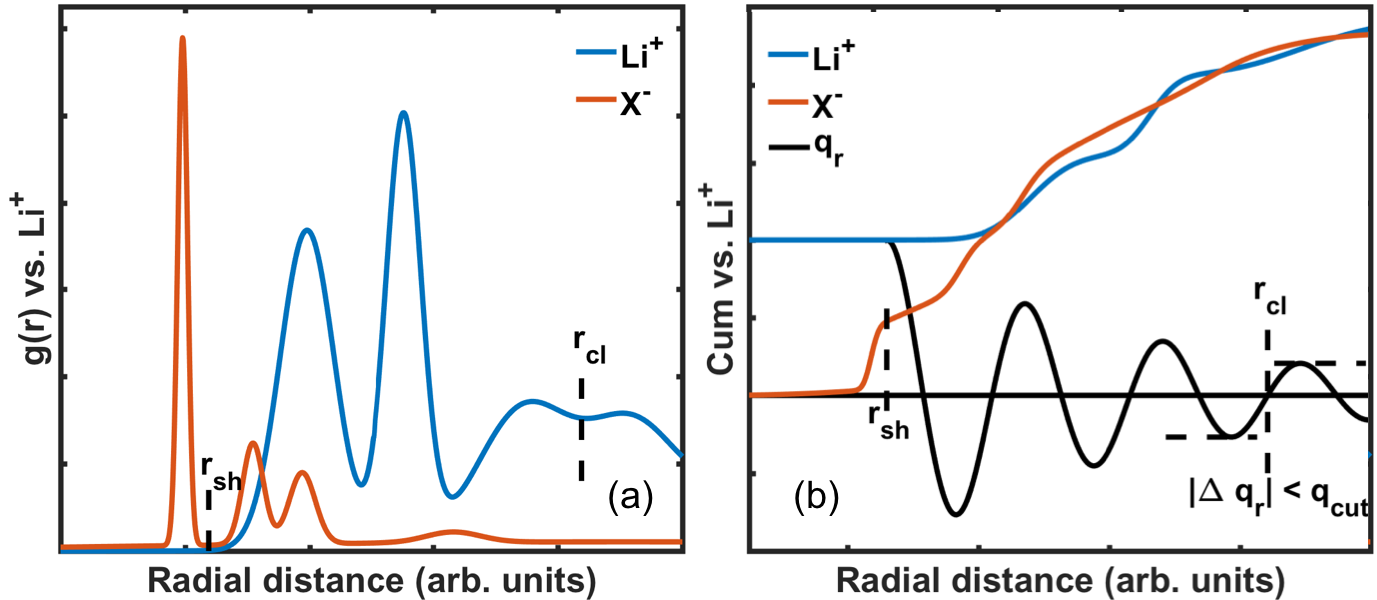}
\caption{Schematic showing typically observed charge variation in an ion pair as a function of radial distance. $r_c$ is the selected cutoff distance defined by the fluctuation point at which the fluctuation width $|\Delta q_{r}|$ is $<$ 0.5. }
\label{fig:scheme}
\end{figure}

To quantify these ideas based on the condition of charge neutrality, i.e.  $|q_{r}|$ $\sim$ 0, we define the radial cutoff limit as $r_{cl}$.  Instead of the actual physical size of the ion pair, $r_{cl}$ can be understood as the radial distance beyond which all charge fluctuations can be assigned as tail end fluctuations outside of the limits of the cluster under consideration. In Fig.~\ref{fig:scheme}, a typical RDF and the corresponding fluctuations in the associated charge curve are depicted schematically as a function of the radial distance $r$ with respect to a reference Li$^+$ ion. As can be observed, such points of charge neutrality usually occur on the charge curve as inflection points between the oppositely charged peaks, indicating the expected ordering where the Li$^+$ ion is first surrounded by counterions which are followed by compensating Li$^+$ ions. The same can be corroborated exactly from the RDF plots in Fig.~\ref{fig:rdfset} for each case. Based on these definitions, we propose the cutoff limit of charge neutrality as the outermost inflection or charge neutral point on the charge curve beyond which the amplitude of charge fluctuation or simply the fluctuation width $|\Delta q_{r}|$ is $<$ $q_{cut}$ = 0.5.

\begin{figure}[!ht]
\centering
\captionsetup{format=plain}
\includegraphics[width=0.8\textwidth]{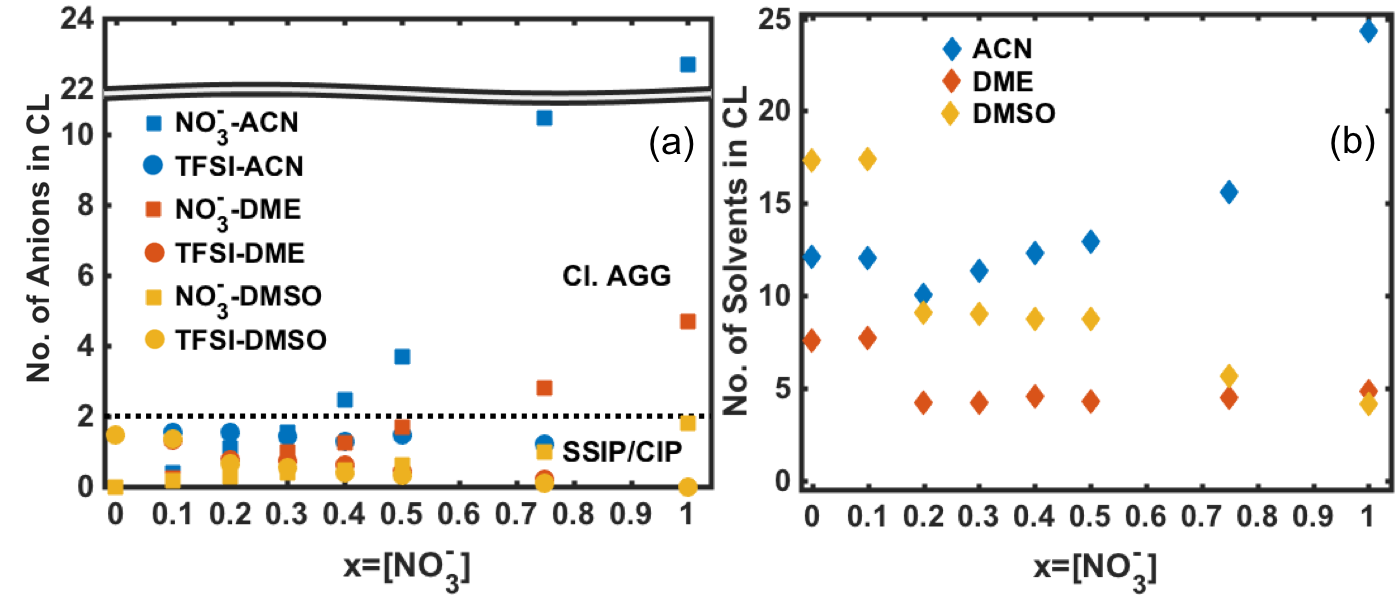}
\caption{(a) No. of Li$^+$ ions contained within a charge neutral ion pair and (b) Cutoff limit $r_{c}$ for defining the limits of the charge neutral ion pair as a function of electrolyte composition.}
\label{fig:CLs}
\end{figure}

The criterion of $q_{cut}$ = 0.5 was chosen after careful analysis of all cumulative and charge plots for all the salt-solvent combinations because the fluctuations were found to dampen quite fast with respect to the distance thereafter, as evidenced in the charge plots the Supplementary Information. The value of $q_{r}$ at $r_{cl}$ is often close to 0 but not exactly 0, as shown schematically in the ideal case in Fig.~\ref{fig:scheme}, implying that there will naturally be some errors associated within this scheme. Subject to these requirements and assumptions, we propose that the number of counterions contained within the charge neutral cluster defined by cutoff limit $r_{cl}$ will decide the degree of ion pairing on a more realistic and continuous scale. It can be intuitively ascertained that the presence of more than 2 counterions, and thus, ion pairs, within the cutoff limit would imply formation of clustered aggregates (Cl. AGG) that also include infused solvent molecules. A value below 2 would imply formation of SSIP or CIP, the distinction between which can be inferred more precisely from the conventional CNs. Besides these, this proposed metric reveals the amount of solvent infused in the clusters. Having established these criteria, which are independent from the choice of salt or solvent in the system, we plot the number of anions and solvents contained within the cutoff limit of the charge-neutral aggregation in Fig.~\ref{fig:CLs}a and Fig.~\ref{fig:CLs}b, respectively.

Using these values from the analysis it is easy to ascertain the extent of clustering leading to aggregation for the simulated salt and solvent combinations. In 1 M TFSI, the weakly associating anions form SSIP/CIPS, as evident from the values $<$ 1 as well as the snapshots. Formation of clustered AGGs takes place at starting at [NO$^-_3$] $>=$ 0.4 M  for the weakly donating. For the strongly donating DMSO, AGGs would probably form only at [NO$_3^-$] $>$ 1 M. For the moderately donating solvent DME, AGGs begin to form only at [NO$_3^-$] $>=$ 0.75 M. While it is clear that strongly associating anions will have a higher degree of aggregation, they cannot exert their effect beyond a certain limit in high DN solvents, as evident from the case of DMSO.

The number of solvents in all cases shows a unique feature where a sharp drop in the number of solvents in the cluster is observed  in every case for 0.1 M $<=$ [NO$^-_3$] $<=$ 0.2 M. This is indicative of a structural transition from a TFSI dominated solvation shell to a NO$^-_3$ dominated solvation shell, in which the number of solvents are fewer as expected. The number of solvents with increasing NO$^-_3$ concentration however, varies quite differently. While the number of ACN molecules increases to as much as the number of counterions, the number of DMSO molecules falls off. This is directly correlated to the ascertained cutoff limits for each of the cases, $r_{cl}$ = 0.910 nm (ACN), $r_{cl}$ = 0.628 nm (DME) and $r_{cl}$ = 0.504 nm (DMSO) at [NO$^-_3$]= 1 M, in which case the immediate solvation shell is dominated by the strongly associating counterion. From these analyses, an estimation of the extent of ion pair formation on a much more continuous scale is obtained, and the same can also be qualitatively correlated to the visualizations in Fig.~\ref{fig:boxset}. The cutoff limits for all other cases are tabulated in the Supplementary Information. 

\begin{figure}[!ht]
\centering
\captionsetup{format=plain}
\includegraphics[width=0.8\textwidth]{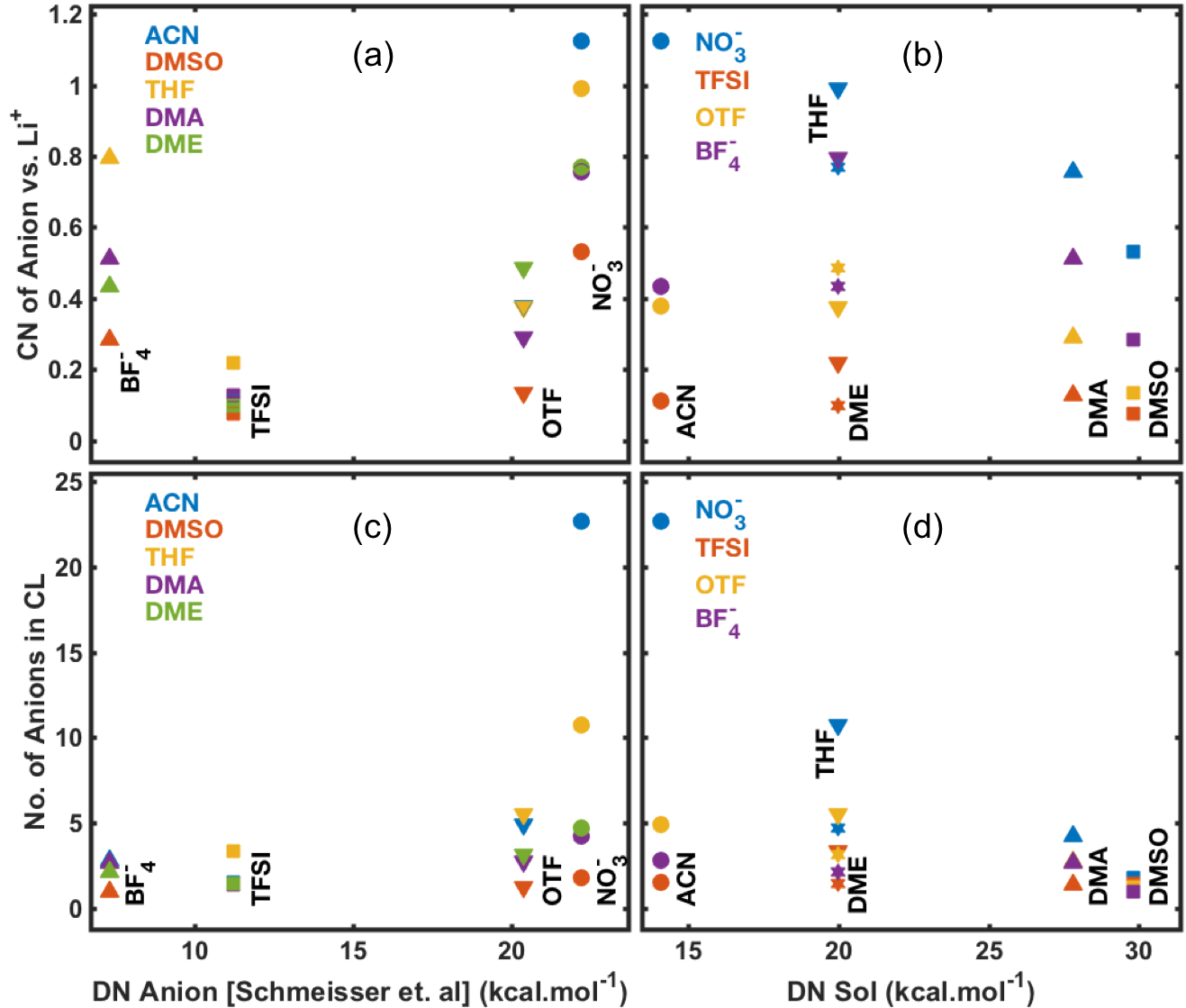}
\caption{Comparison of the two metric discussed in the current work with DNs of the anion and solvent, which are descriptors of electron donicity and ionic association. (a) \textit{top left} CN of anion (vs. Li$^+$) vs. DN of anion ((a) \textit{top right} CN of anion (vs. Li$^+$) vs. DN of solvent (c) \textit{bottom left} Number of anions in the charge neutral cluster vs. DN of anion (d) \textit{bottom right} Number of anions in the charge neutral cluster vs. DN of solvent.}
\label{fig:DNs}
\end{figure}

Finally, we interpret these estimations with respect to a commonly employed descriptor of electron donicity - the Gutmann donor number (DN). As can be seen in Fig.~\ref{fig:DNs}, we plot the solvent and anion donor numbers with respect to the two main metrics discussed in the work, CNs and number of anions in the cluster (CL). In Fig.~\ref{fig:DNs}a and Fig.~\ref{fig:DNs}b, the CNs of the anion vs. Li$^+$ in the first solvation shell are described with respect to the anion's and solvent's DNs, respectively. As can be seen, there doesn't exist a well defined correlation between the DNs and the CNs, where despite being the weakest donating anion, BF$^-_4$ anion has much higher CN in the solvation shell compared to TFSI and OTF. The trends in the CNs with respect to anions is more representative of the often discussed ionic association strength of salt anions~\cite{Henderson2006,HendersonSeo2012a,HendersonSeo2012b}, which is known to follow the order TFSI $<$ BF$^-_4$ $\sim$ OTF $<$ NO$^-_3$. An anion of higher DN is expected to dominate in presence around the Li$^+$ ion, however, the coordination behavior in immediate solvation shell is subject to size and steric effects, as well as salt solvent compatibility. The CN of anions are found to have a much more well defined decreasing trend with respect to the solvents.

In comparison, the number of anions in the clusters as estimated by the metric proposed in this work show much better proportionality with respect to DNs of both the anion and solvent, as can be seen in Fig.~\ref{fig:DNs}c and Fig.~\ref{fig:DNs}d, respectively. This behavior can be rationalized on the grounds that first order descriptors are more representative of larger scale behavior and are not apt for direct comparison with solvation phenomena in the immediate solvation shell. For the considered solvent:Li ratios, as mentioned in the Supplementary Information, our results show good agreement with respect to solubility and aggregation behavior observed in all the anions for ACN~\cite{HendersonSeo2012a,HendersonSeo2012b} and DME~\cite{Henderson2006}. Although clear trends can be inferred from the analysis of coordination numbers as well, the estimation of extent of ion-pairing and clustered aggregation is not directly possible from them.

The application of this scheme is independent of the level of theory or the accuracy of force fields used for simulating the molecular dynamics, although, it must be noted carefully that the classical MD simulations presented here have been performed with non-polarizable forcefields which generally tend to overestimate CNs~\cite{RodeAzam2009}. In comparison to first principles methods, classical MD simulations are very inexpensive and can reveal trends on the basis of which electrolyte components can be analyzed and compared. Besides, this scheme can be applied to both dilute and concentrated electrolytes.

\section{Summary and Conclusions}

In the present work, classical MD simulations were employed to simulate highly concentrated 1 M Li salts in various nonaqueous solvents, both spanning over a wide range of electron doncity, polarizability and size. As a test case, 1 M mixtures of LiNO$^-_3$ + LiTFSI with varying NO$^-_3$:TFSI ratios were also simulated to understand change in solvation structure transitioning from a large weakly donating TFSI anion to a small strongly associating NO$^-_3$ anion. RDFs of each of the systems were analyzed to understand the ordering of anions and solvents around Li$^+$ ions and CNs were estimated by analyzing the cumulative distributions obtained from the integration of the RDFs.  The evaluated CNs were further employed to differentiate and predict the type of ion pairing for the test cases in ACN, DME and DMSO. It was argued that based on the estimated CNs, only 1 M NO$^-_3$, which is the most strongly associating salt considered in the work, would form CIPs in ACN, which is the most weakly donating solvent considered in this work.

In order to resolve this discrepancy, a new metric based on charge neutrality of ionic clusters was detailed. Considering the fact that within some nominal deviation all ionic clusters will essentially have net null charge, a new cutoff limit was devised on the basis of which new estimates for the number of anions, and hence ion pairs, in these charge neutral clusters were obtained. The numbers of ion pairs in such clusters were further used to classify and understand ion pairing categories and thus explore aggregation in a more generalized way which allowed inclusion of solvent molecules in those clustered aggregates. These estimates were then pitted against the donor numbers of both the solvent and the salt anions to show that CNs of anions are rather directly related to commonly understood association strength of anions rather than their donor numbers, which were found to be better correlated to the longer length scale estimated number of ion pairs in charge neutral clusters.

The present work offers a new strategy to understand solubility and ion pairing which can potentially be combined with current efforts to identify novel, stable electrolytes, including those in which organic molecules are entirely removed from the electrolyte~\cite{Addison2016}, to develop an electrolyte that could enable high-energy, long-life Li-O$_2$ batteries. Using these descriptions of ion pairing to model and calculate solvation thermodynamics using cluster models will be the focus of out next research efforts.

\newpage

\begin{suppinfo}
\begin{itemize}
\item A. Figure S1: Plots for normalized radial distribution function, cumulative values and radially contained ionic charge for all solvent-anion combination considered in the present study.
\item B. Table S1: Cumulative numbers of Li$^+$ ions, solvent molecules and anions in the ionic cluster (CL) defined by the cutoff limit $r_{cl}$ for 1 M mixtures of NO$_3^-$+TFSI in solvents ACN, DME and DMSO.
\item C. Table S2: Coordination numbers (CN), cumulative numbers of Li$^+$ ions, solvent molecules and anions in the ionic cluster (CL) defined by the cutoff limit $r_{cl}$ for all salt anion-solvent combinations considered in this work.
\item D. GROMACS forcefield parameters for all considered solvents.
\item E. GROMACS forcefield parameters for all considered salt-anions.
\end{itemize}
\end{suppinfo}

\begin{acknowledgement}
A.K. and H.P. acknowledge funding by the Deutsche Forschungsgemeinschaft (DFG) within the graduate school mobilEM. A.K. also acknowledges the grant of computational time for the project by the J\"{u}lich Aachen Research Alliance (JARA) via Project No. 99000538(JHPC4401). A.K. and V.V. acknowledge support from Convergent Aeronautics Solutions (CAS) project under the NASA Aeronautics Research Mission Directorate.
\end{acknowledgement}

\bibliography{bibsources}

\providecommand{\noopsort}[1]{}\providecommand{\singleletter}[1]{#1}%
\providecommand{\latin}[1]{#1}
\makeatletter
\providecommand{\doi}
  {\begingroup\let\do\@makeother\dospecials
  \catcode`\{=1 \catcode`\}=2 \doi@aux}
\providecommand{\doi@aux}[1]{\endgroup\texttt{#1}}
\makeatother
\providecommand*\mcitethebibliography{\thebibliography}
\csname @ifundefined\endcsname{endmcitethebibliography}
  {\let\endmcitethebibliography\endthebibliography}{}
\begin{mcitethebibliography}{46}
\providecommand*\natexlab[1]{#1}
\providecommand*\mciteSetBstSublistMode[1]{}
\providecommand*\mciteSetBstMaxWidthForm[2]{}
\providecommand*\mciteBstWouldAddEndPuncttrue
  {\def\EndOfBibitem{\unskip.}}
\providecommand*\mciteBstWouldAddEndPunctfalse
  {\let\EndOfBibitem\relax}
\providecommand*\mciteSetBstMidEndSepPunct[3]{}
\providecommand*\mciteSetBstSublistLabelBeginEnd[3]{}
\providecommand*\EndOfBibitem{}
\mciteSetBstSublistMode{f}
\mciteSetBstMaxWidthForm{subitem}{(\alph{mcitesubitemcount})}
\mciteSetBstSublistLabelBeginEnd
  {\mcitemaxwidthsubitemform\space}
  {\relax}
  {\relax}

\bibitem[{A. C. Luntz and B. D McCloskey}(2014)]{McCloskeyLuntz2014}
{A. C. Luntz and B. D McCloskey}, {Nonaqueous Li-Air Batteries: A Status
  Report}. \emph{Chem. Rev.} \textbf{2014}, \emph{114}, 11721--11750\relax
\mciteBstWouldAddEndPuncttrue
\mciteSetBstMidEndSepPunct{\mcitedefaultmidpunct}
{\mcitedefaultendpunct}{\mcitedefaultseppunct}\relax
\EndOfBibitem
\bibitem[{O. Sapunkov and V. Pande and A. Khetan and C. Choomwattana and V.
  Viswanathan}(2015)]{Oleg2015}
{O. Sapunkov and V. Pande and A. Khetan and C. Choomwattana and V.
  Viswanathan}, {{Quantifying the Promise of `Beyond' Li-Ion Batteries}}.
  \emph{Translational Materials Research} \textbf{2015}, \emph{2}, 045002\relax
\mciteBstWouldAddEndPuncttrue
\mciteSetBstMidEndSepPunct{\mcitedefaultmidpunct}
{\mcitedefaultendpunct}{\mcitedefaultseppunct}\relax
\EndOfBibitem
\bibitem[{D. Aurbach and B. D. McCloskey and L. F. Nazar and P. G.
  Bruce}(2016)]{BruceAurbach2016}
{D. Aurbach and B. D. McCloskey and L. F. Nazar and P. G. Bruce}, {{Advances in
  Understanding Mechanisms Underpinning Lithium--Air Batteries}}. \emph{Nat.
  Energy} \textbf{2016}, \emph{1}, 16128\relax
\mciteBstWouldAddEndPuncttrue
\mciteSetBstMidEndSepPunct{\mcitedefaultmidpunct}
{\mcitedefaultendpunct}{\mcitedefaultseppunct}\relax
\EndOfBibitem
\bibitem[{M. J. Trahan and I. Gunasekara and S. Mukerjee and E. J. Plichta and
  M. A. Hendrickson and K. M. Abraham}(2014)]{Trahan2014}
{M. J. Trahan and I. Gunasekara and S. Mukerjee and E. J. Plichta and M. A.
  Hendrickson and K. M. Abraham}, {{Solvent-Coupled Catalysis of the Oxygen
  Electrode Reactions in Lithium-Air Batteries}}. \emph{J. Electrochem. Soc.}
  \textbf{2014}, \emph{161}, A1706--A1715\relax
\mciteBstWouldAddEndPuncttrue
\mciteSetBstMidEndSepPunct{\mcitedefaultmidpunct}
{\mcitedefaultendpunct}{\mcitedefaultseppunct}\relax
\EndOfBibitem
\bibitem[{L. Johnson and C. Li and Z. Liu and Y. Chen and S. A. Freunberger and
  P. Ashok and B. B. Praveen and K. Dholakia and J. Tarascon and P. G.
  Bruce}(2014)]{BruceJohnson2014}
{L. Johnson and C. Li and Z. Liu and Y. Chen and S. A. Freunberger and P. Ashok
  and B. B. Praveen and K. Dholakia and J. Tarascon and P. G. Bruce}, {{The
  Role of LiO$_2$ Solubility in O$_2$ Reduction in Aprotic Solvents and its
  Consequences for Li-O$_2$ Batteries}}. \emph{Nat. Chem.} \textbf{2014},
  \emph{6}, 1091--1099\relax
\mciteBstWouldAddEndPuncttrue
\mciteSetBstMidEndSepPunct{\mcitedefaultmidpunct}
{\mcitedefaultendpunct}{\mcitedefaultseppunct}\relax
\EndOfBibitem
\bibitem[{A. Khetan and A. C. Luntz and V. Viswanathan}(2015)]{Khetan:2015aa}
{A. Khetan and A. C. Luntz and V. Viswanathan}, {{Trade-offs in Capacity and
  Rechargeability in Nonaqueous Li-O$_2$ Batteries: Solution-Driven Growth
  versus Nucleophilic Stability}}. \emph{J. Phys. Chem. Lett.} \textbf{2015},
  \emph{6}, 1254--1259\relax
\mciteBstWouldAddEndPuncttrue
\mciteSetBstMidEndSepPunct{\mcitedefaultmidpunct}
{\mcitedefaultendpunct}{\mcitedefaultseppunct}\relax
\EndOfBibitem
\bibitem[{A. Khetan and H. Pitsch and V. Viswanathan}(2014)]{Khetan:2014ab}
{A. Khetan and H. Pitsch and V. Viswanathan}, {{Solvent Degradation in
  Nonaqueous Li-O$_2$ Batteries: Oxidative Stability versus H-Abstraction}}.
  \emph{J. Phys. Chem. Lett.} \textbf{2014}, \emph{5}, 2419--2424\relax
\mciteBstWouldAddEndPuncttrue
\mciteSetBstMidEndSepPunct{\mcitedefaultmidpunct}
{\mcitedefaultendpunct}{\mcitedefaultseppunct}\relax
\EndOfBibitem
\bibitem[{A. Khetan and H. Pitsch and V. Viswanathan}(2014)]{Khetan:2014aa}
{A. Khetan and H. Pitsch and V. Viswanathan}, {Identifying Descriptors for
  Solvent Stability in Nonaqueous Li-O$_2$ Batteries}. \emph{J. Phys. Chem.
  Lett.} \textbf{2014}, \emph{5}, 1318--1323\relax
\mciteBstWouldAddEndPuncttrue
\mciteSetBstMidEndSepPunct{\mcitedefaultmidpunct}
{\mcitedefaultendpunct}{\mcitedefaultseppunct}\relax
\EndOfBibitem
\bibitem[{E. Nasybulin and W. Xu and M. H. Engelhard and Z. Nie and S. D.
  Burton and L. Cosimbescu and M. E. Gross and J. Zhang}(2013)]{Nasybulin2013}
{E. Nasybulin and W. Xu and M. H. Engelhard and Z. Nie and S. D. Burton and L.
  Cosimbescu and M. E. Gross and J. Zhang}, {{Effects of Electrolyte Salts on
  the Performance of Li-O$_2$ Batteries}}. \emph{J. Phys. Chem. C}
  \textbf{2013}, \emph{117}, 2635--2645\relax
\mciteBstWouldAddEndPuncttrue
\mciteSetBstMidEndSepPunct{\mcitedefaultmidpunct}
{\mcitedefaultendpunct}{\mcitedefaultseppunct}\relax
\EndOfBibitem
\bibitem[{I. Gunasekara and S. Mukerjee and E. J. Plichta and M. A. Hendrickson
  and K. M. Abraham}(2015)]{Gunasekara2015}
{I. Gunasekara and S. Mukerjee and E. J. Plichta and M. A. Hendrickson and K.
  M. Abraham}, {{A Study of the Influence of Lithium Salt Anions on Oxygen
  Reduction Reactions in Li-Air Batteries}}. \emph{J. Electrochem. Soc.}
  \textbf{2015}, \emph{162}, A1055--A1066\relax
\mciteBstWouldAddEndPuncttrue
\mciteSetBstMidEndSepPunct{\mcitedefaultmidpunct}
{\mcitedefaultendpunct}{\mcitedefaultseppunct}\relax
\EndOfBibitem
\bibitem[{D. Sharon and D. Hirsberg and M. Salama and M. Afri and A. Frimer and
  M. Noked and W. Kwak and Y. Sun and D. Aurbach}(2016)]{AurbachSharon2016}
{D. Sharon and D. Hirsberg and M. Salama and M. Afri and A. Frimer and M. Noked
  and W. Kwak and Y. Sun and D. Aurbach}, {{Mechanistic Role of Li$^+$
  Dissociation Level in Aprotic Li-O$_2$ Battery}}. \emph{ACS Appl. Mater.
  Interfaces} \textbf{2016}, \emph{8}, 5300--5307\relax
\mciteBstWouldAddEndPuncttrue
\mciteSetBstMidEndSepPunct{\mcitedefaultmidpunct}
{\mcitedefaultendpunct}{\mcitedefaultseppunct}\relax
\EndOfBibitem
\bibitem[{N. Aetukuri and B. D. McCloskey and L. E. Krupp and V. Viswanathan
  and A. C. Luntz}(2015)]{Aetukuri2015}
{N. Aetukuri and B. D. McCloskey and L. E. Krupp and V. Viswanathan and A. C.
  Luntz}, {{Solvating Additives Drive Solution-Mediated Electrochemistry and
  Enhance Toroid Growth in Non-aqueous Li-O$_2$ Batteries}}. \emph{Nat. Chem.}
  \textbf{2015}, \emph{7}, 50--56\relax
\mciteBstWouldAddEndPuncttrue
\mciteSetBstMidEndSepPunct{\mcitedefaultmidpunct}
{\mcitedefaultendpunct}{\mcitedefaultseppunct}\relax
\EndOfBibitem
\bibitem[{X. Gao and Y. Chen and L. Johnson and P. G.
  Bruce}(2016)]{BruceGao2016}
{X. Gao and Y. Chen and L. Johnson and P. G. Bruce}, {{Promoting Solution Phase
  Discharge in Li-O$_2$ Batteries Containing Weakly Solvating Electrolyte
  Solutions}}. \emph{Nat. Mater.} \textbf{2016}, \emph{15}, 882--888\relax
\mciteBstWouldAddEndPuncttrue
\mciteSetBstMidEndSepPunct{\mcitedefaultmidpunct}
{\mcitedefaultendpunct}{\mcitedefaultseppunct}\relax
\EndOfBibitem
\bibitem[{C. M. Burke and V. Pande and A. Khetan and V. Viswanathan and B. D.
  McCloskey}(2015)]{Burke2015}
{C. M. Burke and V. Pande and A. Khetan and V. Viswanathan and B. D.
  McCloskey}, {{Enhancing Electrochemical Intermediate Solvation through
  Electrolyte Anion Selection to Increase Nonaqueous Li-O$_2$ Battery
  Capacity}}. \emph{Proc. Natl. Acad. Sci. U.S.A} \textbf{2015}, \emph{112},
  9293--9298\relax
\mciteBstWouldAddEndPuncttrue
\mciteSetBstMidEndSepPunct{\mcitedefaultmidpunct}
{\mcitedefaultendpunct}{\mcitedefaultseppunct}\relax
\EndOfBibitem
\bibitem[{W. Walker and V. Giordani and J. Uddin and V. S. Bryantsev and G. V.
  Chase and D. Addison}(2013)]{Walker:2013aa}
{W. Walker and V. Giordani and J. Uddin and V. S. Bryantsev and G. V. Chase and
  D. Addison}, {A Rechargeable Li-O$_2$ Battery Using a Lithium
  Nitrate/N,N-Dimethylacetamide Electrolyte}. \emph{J. Am. Chem. Soc.}
  \textbf{2013}, \emph{135}, 2076--2079\relax
\mciteBstWouldAddEndPuncttrue
\mciteSetBstMidEndSepPunct{\mcitedefaultmidpunct}
{\mcitedefaultendpunct}{\mcitedefaultseppunct}\relax
\EndOfBibitem
\bibitem[{D. Sharon and D. Hirsberg and M. Afri and F. Chesneau and R. Lavi and
  A. Frimer and Y. Sun and D. Aurbach}(2015)]{AurbachSharon2015}
{D. Sharon and D. Hirsberg and M. Afri and F. Chesneau and R. Lavi and A.
  Frimer and Y. Sun and D. Aurbach}, {{Catalytic Behavior of Lithium Nitrate in
  Li-O$_2$ Cells}}. \emph{ACS Appl. Mater. Interfaces} \textbf{2015}, \emph{7},
  16590--16600\relax
\mciteBstWouldAddEndPuncttrue
\mciteSetBstMidEndSepPunct{\mcitedefaultmidpunct}
{\mcitedefaultendpunct}{\mcitedefaultseppunct}\relax
\EndOfBibitem
\bibitem[{S. J. Kang and T. Mori and S. Narizuka and W. Wilcke and H.
  Kim}(2014)]{KimKang2014}
{S. J. Kang and T. Mori and S. Narizuka and W. Wilcke and H. Kim},
  {{Deactivation of Carbon Electrode for Elimination of Carbon Dioxide
  Evolution from Rechargeable Lithium--oxygen Cells}}. \emph{Nat. Commun.}
  \textbf{2014}, \emph{5}\relax
\mciteBstWouldAddEndPuncttrue
\mciteSetBstMidEndSepPunct{\mcitedefaultmidpunct}
{\mcitedefaultendpunct}{\mcitedefaultseppunct}\relax
\EndOfBibitem
\bibitem[{M. Iliksu and A. Khetan and S. Yang and U. Simon and H. Pitsch and D.
  U. Sauer}(2017)]{SauerIliksu2017}
{M. Iliksu and A. Khetan and S. Yang and U. Simon and H. Pitsch and D. U.
  Sauer}, {{Elucidation and Comparison of the Effect of LiTFSI and LiNO$_3$
  Salts on Discharge Chemistry in Nonaqueous Li-O$_2$ Batteries}}. \emph{ACS
  Appl. Mater. Interfaces} \textbf{2017}, \emph{9}, 19319--19325\relax
\mciteBstWouldAddEndPuncttrue
\mciteSetBstMidEndSepPunct{\mcitedefaultmidpunct}
{\mcitedefaultendpunct}{\mcitedefaultseppunct}\relax
\EndOfBibitem
\bibitem[{Y. Marcus and G. Hefter}(2006)]{Marcus2006}
{Y. Marcus and G. Hefter}, Ion Pairing. \emph{Chem. Rev.} \textbf{2006},
  \emph{106}, 4585--4621\relax
\mciteBstWouldAddEndPuncttrue
\mciteSetBstMidEndSepPunct{\mcitedefaultmidpunct}
{\mcitedefaultendpunct}{\mcitedefaultseppunct}\relax
\EndOfBibitem
\bibitem[{P. Du and J. Lu and K. C. Lau and X. Luo and J. Bareno and X. Zhang
  and Y. Ren and Z. Zhang and L. A. Curtiss and Y. Sun and K.
  Amine}(2013)]{AmineDu2013}
{P. Du and J. Lu and K. C. Lau and X. Luo and J. Bareno and X. Zhang and Y. Ren
  and Z. Zhang and L. A. Curtiss and Y. Sun and K. Amine}, {{Compatibility of
  Lithium Salts with Solvent of the Non-aqueous Electrolyte in Li-O$_2$
  Batteries}}. \emph{Phys. Chem. Chem. Phys.} \textbf{2013}, \emph{15},
  5572--5581\relax
\mciteBstWouldAddEndPuncttrue
\mciteSetBstMidEndSepPunct{\mcitedefaultmidpunct}
{\mcitedefaultendpunct}{\mcitedefaultseppunct}\relax
\EndOfBibitem
\bibitem[{N. N. Rajput and X. Qu and N. Sa and A. K. Burrell and K. A.
  Persson}(2015)]{PerssonRajput2015}
{N. N. Rajput and X. Qu and N. Sa and A. K. Burrell and K. A. Persson}, {{The
  Coupling between Stability and Ion Pair Formation in Magnesium Electrolytes
  from First-Principles Quantum Mechanics and Classical Molecular Dynamics}}.
  \emph{J. Am. Chem. Soc.} \textbf{2015}, \emph{137}, 3411--3420\relax
\mciteBstWouldAddEndPuncttrue
\mciteSetBstMidEndSepPunct{\mcitedefaultmidpunct}
{\mcitedefaultendpunct}{\mcitedefaultseppunct}\relax
\EndOfBibitem
\bibitem[{V. S. Bryantsev and J. Uddin and V. Giordani and W. Walker and D.
  Addison and G. V. Chase}(2013)]{Bryantsev:2013ab}
{V. S. Bryantsev and J. Uddin and V. Giordani and W. Walker and D. Addison and
  G. V. Chase}, {The Identification of Stable Solvents for Nonaqueous
  Rechargeable Li-Air Batteries}. \emph{J. Electrochem. Soc.} \textbf{2013},
  \emph{160}, A160--A171\relax
\mciteBstWouldAddEndPuncttrue
\mciteSetBstMidEndSepPunct{\mcitedefaultmidpunct}
{\mcitedefaultendpunct}{\mcitedefaultseppunct}\relax
\EndOfBibitem
\bibitem[{K. Ueno and K. Yoshida and M. Tsuchiya and N. Tachikawa and K. Dokko
  and M. Watanabe}(2012)]{WatanabeUeno2012}
{K. Ueno and K. Yoshida and M. Tsuchiya and N. Tachikawa and K. Dokko and M.
  Watanabe}, {{Glyme--Lithium Salt Equimolar Molten Mixtures: Concentrated
  Solutions or Solvate Ionic Liquids?}} \emph{J. Phys. Chem. B} \textbf{2012},
  \emph{116}, 11323--11331\relax
\mciteBstWouldAddEndPuncttrue
\mciteSetBstMidEndSepPunct{\mcitedefaultmidpunct}
{\mcitedefaultendpunct}{\mcitedefaultseppunct}\relax
\EndOfBibitem
\bibitem[{V. Gutmann}(1976)]{Gutmann:1976aa}
{V. Gutmann}, {{Solvent Effects on the Reactivities of Organometallic
  Compounds}}. \emph{Coord. Chem. Rev.} \textbf{1976}, \emph{18}, 225 --
  255\relax
\mciteBstWouldAddEndPuncttrue
\mciteSetBstMidEndSepPunct{\mcitedefaultmidpunct}
{\mcitedefaultendpunct}{\mcitedefaultseppunct}\relax
\EndOfBibitem
\bibitem[{U. Mayer}(1979)]{Mayer:1979aa}
{U. Mayer}, {{A Semiempirical Model for the Description of Solvent Effects on
  Chemical Reactions}}. \emph{Pure Appl. Chem.} \textbf{1979}, \emph{51},
  1697--1712\relax
\mciteBstWouldAddEndPuncttrue
\mciteSetBstMidEndSepPunct{\mcitedefaultmidpunct}
{\mcitedefaultendpunct}{\mcitedefaultseppunct}\relax
\EndOfBibitem
\bibitem[{M. Schmeisser and P. Illner and R. Puchta and A. Zahl and R.
  vanEldik}(2012)]{Schmeisser:2012}
{M. Schmeisser and P. Illner and R. Puchta and A. Zahl and R. vanEldik},
  {{Gutmann Donor and Acceptor Numbers for Ionic Liquids}}. \emph{Chemistry}
  \textbf{2012}, \emph{18}, 10969--10982\relax
\mciteBstWouldAddEndPuncttrue
\mciteSetBstMidEndSepPunct{\mcitedefaultmidpunct}
{\mcitedefaultendpunct}{\mcitedefaultseppunct}\relax
\EndOfBibitem
\bibitem[{D. M. Seo and O. Borodin and S. Han and Q. Ly and P. D. Boyle and W.
  A. Henderson}(2012)]{HendersonSeo2012a}
{D. M. Seo and O. Borodin and S. Han and Q. Ly and P. D. Boyle and W. A.
  Henderson}, {{Electrolyte Solvation and Ionic Association}}. \emph{J.
  Electrochem. Soc.} \textbf{2012}, \emph{159}, A553--A565\relax
\mciteBstWouldAddEndPuncttrue
\mciteSetBstMidEndSepPunct{\mcitedefaultmidpunct}
{\mcitedefaultendpunct}{\mcitedefaultseppunct}\relax
\EndOfBibitem
\bibitem[{D. M. Seo and O. Borodin and S. Han and P. D. Boyle and W. A.
  Henderson}(2012)]{HendersonSeo2012b}
{D. M. Seo and O. Borodin and S. Han and P. D. Boyle and W. A. Henderson},
  {{Electrolyte Solvation and Ionic Association II. Acetonitrile-Lithium Salt
  Mixtures: Highly Dissociated Salts}}. \emph{J. Electrochem. Soc.}
  \textbf{2012}, \emph{159}, A1489--A1500\relax
\mciteBstWouldAddEndPuncttrue
\mciteSetBstMidEndSepPunct{\mcitedefaultmidpunct}
{\mcitedefaultendpunct}{\mcitedefaultseppunct}\relax
\EndOfBibitem
\bibitem[{D. M. Seo and O. Borodin and D. Balogh and M. O'Connell and Q. Ly and
  S. Han and S. Passerini and W. A. Henderson}(2013)]{HendersonSeo2013}
{D. M. Seo and O. Borodin and D. Balogh and M. O'Connell and Q. Ly and S. Han
  and S. Passerini and W. A. Henderson}, {{Electrolyte Solvation and Ionic
  Association III. Acetonitrile-Lithium Salt Mixtures--Transport Properties}}.
  \emph{J. Electrochem. Soc.} \textbf{2013}, \emph{160}, A1061--A1070\relax
\mciteBstWouldAddEndPuncttrue
\mciteSetBstMidEndSepPunct{\mcitedefaultmidpunct}
{\mcitedefaultendpunct}{\mcitedefaultseppunct}\relax
\EndOfBibitem
\bibitem[{S. Han and O. Borodin and J. L. Allen and D. M. Seo and D. W. McOwen
  and S. Yun and W. A. Henderson}(2013)]{HendersonHan2013}
{S. Han and O. Borodin and J. L. Allen and D. M. Seo and D. W. McOwen and S.
  Yun and W. A. Henderson}, {{Electrolyte Solvation and Ionic Association: IV.
  Acetonitrile-Lithium Difluoro(oxalato)borate (LiDFOB) Mixtures}}. \emph{J.
  Electrochem. Soc.} \textbf{2013}, \emph{160}, A2100--A2110\relax
\mciteBstWouldAddEndPuncttrue
\mciteSetBstMidEndSepPunct{\mcitedefaultmidpunct}
{\mcitedefaultendpunct}{\mcitedefaultseppunct}\relax
\EndOfBibitem
\bibitem[{S. Han and O. Borodin and D. M. Seo and Z. Zhou and W. A.
  Henderson}(2014)]{HendersonHan2014}
{S. Han and O. Borodin and D. M. Seo and Z. Zhou and W. A. Henderson},
  {{Electrolyte Solvation and Ionic Association: V. Acetonitrile-Lithium
  Bis(fluorosulfonyl)imide (LiFSI) Mixtures}}. \emph{J. Electrochem. Soc.}
  \textbf{2014}, \emph{161}, A2042--A2053\relax
\mciteBstWouldAddEndPuncttrue
\mciteSetBstMidEndSepPunct{\mcitedefaultmidpunct}
{\mcitedefaultendpunct}{\mcitedefaultseppunct}\relax
\EndOfBibitem
\bibitem[{S. Jung and F. F. Canova and K. Akagi}(2016)]{AkagiJung2016}
{S. Jung and F. F. Canova and K. Akagi}, {{Characteristics of Lithium Ions and
  Superoxide Anions in EMI-TFSI and Dimethyl Sulfoxide}}. \emph{J. Phys. Chem.
  A} \textbf{2016}, \emph{120}, 364--371\relax
\mciteBstWouldAddEndPuncttrue
\mciteSetBstMidEndSepPunct{\mcitedefaultmidpunct}
{\mcitedefaultendpunct}{\mcitedefaultseppunct}\relax
\EndOfBibitem
\bibitem[{N. Kumar and J. M. Seminario}(2016)]{SeminarioKumar2016}
{N. Kumar and J. M. Seminario}, {{Lithium-Ion Model Behavior in an Ethylene
  Carbonate Electrolyte Using Molecular Dynamics}}. \emph{J. Phys. Chem. C}
  \textbf{2016}, \emph{120}, 16322--16332\relax
\mciteBstWouldAddEndPuncttrue
\mciteSetBstMidEndSepPunct{\mcitedefaultmidpunct}
{\mcitedefaultendpunct}{\mcitedefaultseppunct}\relax
\EndOfBibitem
\bibitem[{S. H. Lapidus and N. N. Rajput and X. Qu and K. W. Chapman and K. A.
  Persson and P. J. Chupas}(2014)]{ChupasLapidus2014}
{S. H. Lapidus and N. N. Rajput and X. Qu and K. W. Chapman and K. A. Persson
  and P. J. Chupas}, {{Solvation Structure and Energetics of Electrolytes for
  Multivalent Energy Storage}}. \emph{Phys. Chem. Chem. Phys.} \textbf{2014},
  \emph{16}, 21941--21945\relax
\mciteBstWouldAddEndPuncttrue
\mciteSetBstMidEndSepPunct{\mcitedefaultmidpunct}
{\mcitedefaultendpunct}{\mcitedefaultseppunct}\relax
\EndOfBibitem
\bibitem[{S. Tsuzuki and W. Shinoda and M. Matsugami and Y. Umebayashi and K.
  Ueno and T. Mandai and S. Seki and K. Dokko and M.
  Watanabe}(2015)]{WatanabeTsuzuki2015}
{S. Tsuzuki and W. Shinoda and M. Matsugami and Y. Umebayashi and K. Ueno and
  T. Mandai and S. Seki and K. Dokko and M. Watanabe}, {{Structures of
  [Li(glyme)]$^+$ Complexes and their Interactions with Anions in Equimolar
  Mixtures of Glymes and Li[TFSA]: Analysis by Molecular Dynamics
  Simulations}}. \emph{Phys. Chem. Chem. Phys.} \textbf{2015}, \emph{17},
  126--129\relax
\mciteBstWouldAddEndPuncttrue
\mciteSetBstMidEndSepPunct{\mcitedefaultmidpunct}
{\mcitedefaultendpunct}{\mcitedefaultseppunct}\relax
\EndOfBibitem
\bibitem[{S. Saito and H. Watanabe and K. Ueno and T. Mandai and S. Seki and S.
  Tsuzuki and Y. Kameda and K. Dokko and M. Watanabe and Y.
  Umebayashi}(2016)]{WatanabeSaito2016}
{S. Saito and H. Watanabe and K. Ueno and T. Mandai and S. Seki and S. Tsuzuki
  and Y. Kameda and K. Dokko and M. Watanabe and Y. Umebayashi}, {{Li$^+$ Local
  Structure in Hydrofluoroether Diluted Li-Glyme Solvate Ionic Liquid}}.
  \emph{J. Phys. Chem. B} \textbf{2016}, \emph{120}, 3378--3387\relax
\mciteBstWouldAddEndPuncttrue
\mciteSetBstMidEndSepPunct{\mcitedefaultmidpunct}
{\mcitedefaultendpunct}{\mcitedefaultseppunct}\relax
\EndOfBibitem
\bibitem[{N. Kuritz and M. Murat and M. Balaish and Y. Ein-Eli and A.
  Natan}(2016)]{NatanKuritz2016}
{N. Kuritz and M. Murat and M. Balaish and Y. Ein-Eli and A. Natan}, {{PFC and
  Triglyme for Li--Air Batteries: A Molecular Dynamics Study}}. \emph{J. Phys.
  Chem. B} \textbf{2016}, \emph{120}, 3370--3377\relax
\mciteBstWouldAddEndPuncttrue
\mciteSetBstMidEndSepPunct{\mcitedefaultmidpunct}
{\mcitedefaultendpunct}{\mcitedefaultseppunct}\relax
\EndOfBibitem
\bibitem[{W. A. Henderson}(2006)]{Henderson2006}
{W. A. Henderson}, {{Glyme−Lithium Salt Phase Behavior}}. \emph{J. Phys.
  Chem. B} \textbf{2006}, \emph{110}, 13177--13183\relax
\mciteBstWouldAddEndPuncttrue
\mciteSetBstMidEndSepPunct{\mcitedefaultmidpunct}
{\mcitedefaultendpunct}{\mcitedefaultseppunct}\relax
\EndOfBibitem
\bibitem[{M. J. Abraham and T. Murtola and R. Schulz and S. P{\'a}ll and J. C.
  Smith and B. Hess and E. Lindahl}(2015)]{LindahlAbraham2015}
{M. J. Abraham and T. Murtola and R. Schulz and S. P{\'a}ll and J. C. Smith and
  B. Hess and E. Lindahl}, {{GROMACS: High Performance Molecular Simulations
  through Multi-level Parallelism from Laptops to Supercomputers}}.
  \emph{SoftwareX} \textbf{2015}, \emph{1--2}, 19 -- 25\relax
\mciteBstWouldAddEndPuncttrue
\mciteSetBstMidEndSepPunct{\mcitedefaultmidpunct}
{\mcitedefaultendpunct}{\mcitedefaultseppunct}\relax
\EndOfBibitem
\bibitem[{D. van der Spoel and P. J. van Maaren and C.
  Caleman}(2012)]{CalemanSpoel2012}
{D. van der Spoel and P. J. van Maaren and C. Caleman}, {{GROMACS Molecule $\&$
  Liquid Database}}. \emph{Bioinformatics} \textbf{2012}, \emph{28},
  752--753\relax
\mciteBstWouldAddEndPuncttrue
\mciteSetBstMidEndSepPunct{\mcitedefaultmidpunct}
{\mcitedefaultendpunct}{\mcitedefaultseppunct}\relax
\EndOfBibitem
\bibitem[{C. Caleman and P. J. van Maaren and M. Hong and J. S. Hub and L. T.
  Costa and D. van der Spoel}(2012)]{SpoelCaleman2012}
{C. Caleman and P. J. van Maaren and M. Hong and J. S. Hub and L. T. Costa and
  D. van der Spoel}, {{Force Field Benchmark of Organic Liquids: Density,
  Enthalpy of Vaporization, Heat Capacities, Surface Tension, Isothermal
  Compressibility, Volumetric Expansion Coefficient, and Dielectric Constant}}.
  \emph{J. Chem. Theory Comput.} \textbf{2012}, \emph{8}, 61--74\relax
\mciteBstWouldAddEndPuncttrue
\mciteSetBstMidEndSepPunct{\mcitedefaultmidpunct}
{\mcitedefaultendpunct}{\mcitedefaultseppunct}\relax
\EndOfBibitem
\bibitem[{S. V. Sambasivarao and O. Acevedo}(2009)]{AcevedoRao2009}
{S. V. Sambasivarao and O. Acevedo}, {{Development of OPLS-AA Force Field
  Parameters for 68 Unique Ionic Liquids}}. \emph{J. Chem. Theory Comput.}
  \textbf{2009}, \emph{5}, 1038--1050\relax
\mciteBstWouldAddEndPuncttrue
\mciteSetBstMidEndSepPunct{\mcitedefaultmidpunct}
{\mcitedefaultendpunct}{\mcitedefaultseppunct}\relax
\EndOfBibitem
\bibitem[{J. N. Canongia Lopes, Jos{\'e} N. and A. A. H.
  P{\'a}dua}(2004)]{PaduaLopes2004}
{J. N. Canongia Lopes, Jos{\'e} N. and A. A. H. P{\'a}dua}, {{Molecular Force
  Field for Ionic Liquids Composed of Triflate or Bistriflylimide Anions}}.
  \emph{J. Phys. Chem. B} \textbf{2004}, \emph{108}, 16893--16898\relax
\mciteBstWouldAddEndPuncttrue
\mciteSetBstMidEndSepPunct{\mcitedefaultmidpunct}
{\mcitedefaultendpunct}{\mcitedefaultseppunct}\relax
\EndOfBibitem
\bibitem[{S. S. Azam and T. S. Hofer and B. R. Randolf and B. M.
  Rode}(2009)]{RodeAzam2009}
{S. S. Azam and T. S. Hofer and B. R. Randolf and B. M. Rode}, {{Hydration of
  Sodium(I) and Potassium(I) Revisited: A Comparative QM/MM and QMCF MD
  Simulation Study of Weakly Hydrated Ions}}. \emph{J. Phys. Chem. A}
  \textbf{2009}, \emph{113}, 1827--1834\relax
\mciteBstWouldAddEndPuncttrue
\mciteSetBstMidEndSepPunct{\mcitedefaultmidpunct}
{\mcitedefaultendpunct}{\mcitedefaultseppunct}\relax
\EndOfBibitem
\bibitem[{V. Giordani and D. Tozier and H. Tan and C. M. Burke and B. M.
  Gallant and J. Uddin and J. R. Greer and B. D. McCloskey and G. V. Chase and
  D. Addison}(2016)]{Addison2016}
{V. Giordani and D. Tozier and H. Tan and C. M. Burke and B. M. Gallant and J.
  Uddin and J. R. Greer and B. D. McCloskey and G. V. Chase and D. Addison},
  {{A Molten Salt Lithium-Oxygen Battery}}. \emph{J. Am. Chem. Soc.}
  \textbf{2016}, \emph{138}, 2656--2663\relax
\mciteBstWouldAddEndPuncttrue
\mciteSetBstMidEndSepPunct{\mcitedefaultmidpunct}
{\mcitedefaultendpunct}{\mcitedefaultseppunct}\relax
\EndOfBibitem
\end{mcitethebibliography}

\includepdf[pages=1-26]{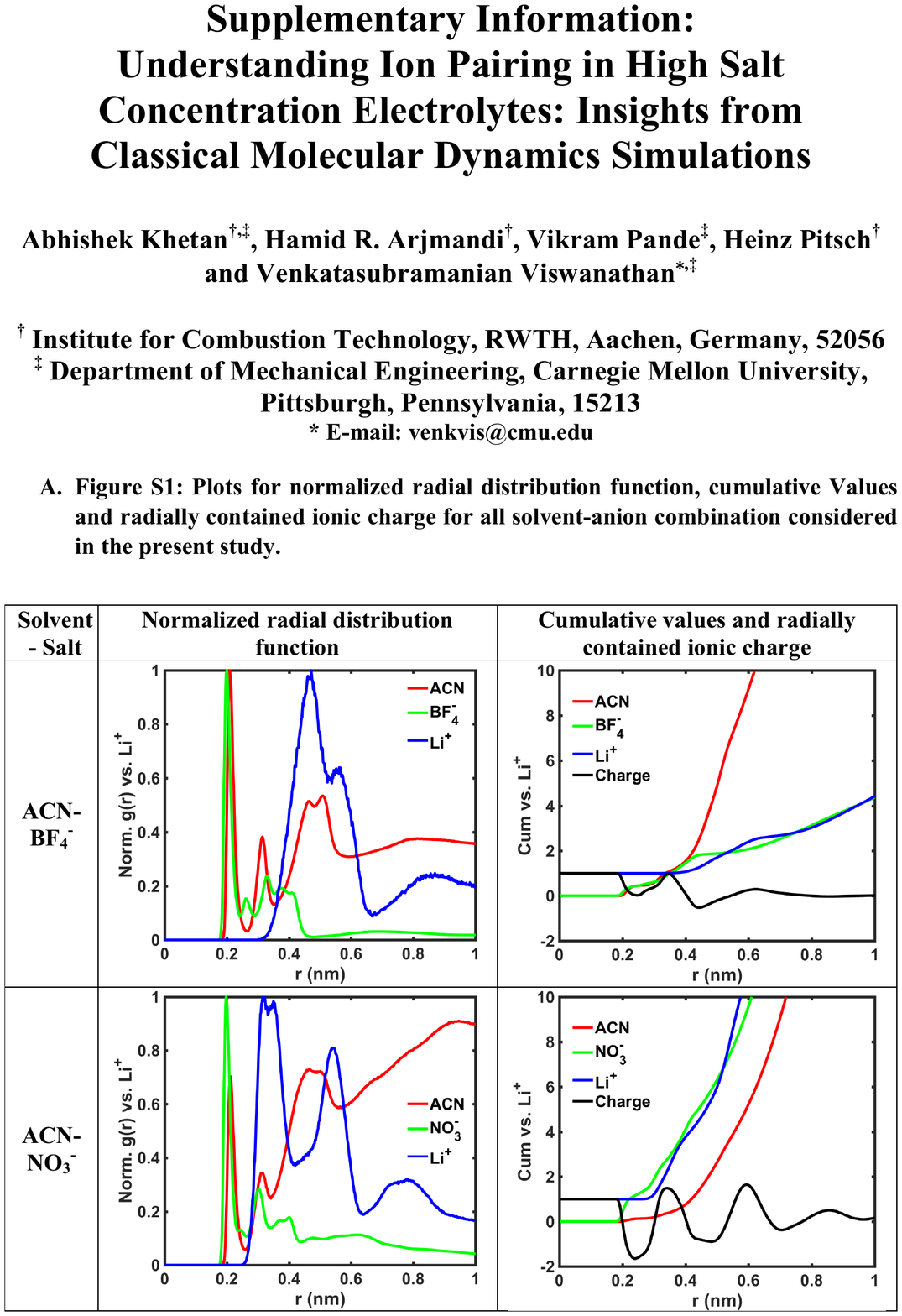}
\end{document}